Correspondence and requests for materials should be addressed to Qing-ming Zhang (qmzhang@ruc.edu.cn)

# Superconductivity at 44 K in K intercalated FeSe system with excess Fe

An-min Zhang,[1] Tian-long Xia, [1] Kai Liu,[1] Wei Tong,[2] Zhao-rong Yang[3] and Qing-ming Zhang*,[1]

[1]Department of Physics, Renmin University of China, Beijing 100872, P. R. China

[2]High Magnetic Field Laboratory, Hefei Institutes of Physical Science, Chinese Academy of Sciences, Hefei 230031,P. R. China

[3]Key Laboratory of Materials Physics, Institute of Solid State Physics, Chinese Academy of Sciences, Hefei 230031, P. R. China

We report here that a new superconducting phase with much higher Tc has been found in K intercalated FeSe compound with excess Fe. We successfully grew crystals by precisely controlling the starting amount of Fe. Besides the superconducting (SC) transition at ~30 K, we observed a sharp drop in resistivity and a kink in susceptibility at 44 K. By combining thermodynamic measurements with electron spin resonance (ESR), we demonstrate that this is a new SC transition. Structural analysis unambiguously reveals two phases coexisting in the crystals, which are responsible respectively for the SC transitions at 30 and 44 K. The structural experiments and first-principles calculations consistently indicate that the 44 K SC phase is close to a 122 structure, but with an unexpectedly large c-axis of 18.10 Å. We further find a novel monotonic dependence of the maximum Tc on the separation of neighbouring FeSe layers.

FeSe-based superconductors show great potential in achieving high transition temperature. The transition temperature of parent compound FeSe is only 8 K,[1] but can be dramatically raised to ~37 K under high pressure.[2] The successful synthesis of $A_xFe_{2-y}Se_2$ (A=K, Rb, Cs, Tl) is an exciting breakthrough. At ambient pressure its maximum Tc can reach 31 K. In the new compound monovalent A ions are intercalated into interstitial sites between FeSe layers.[3] Its superconductivity is considered more likely to originate from Fe-deficient and vacancy-ordered $A_{0.8}Fe_{1.6}Se_2$ phase, while the issue on phase separation has also been raised in the supconductors.[4-9] With a rapid progress in material synthesis, there are already some indications that the maximum Tc under atmosphere pressure may go beyond 31 K in FeSe-based systems. A kink around 40 K in both resistivity and susceptibility was noticed in two earlier studies, which was speculated to be a new SC phase.[11,12] The reason for non-zero resistivity is that SC grains may be too small to form an effective supercurrent network through crystal. However, one may argue that an antiferromagnetic (AF) transition can also cause a similar behavior in resistivity and susceptibility, as observed in FeTe.[13,25] High-pressure experiments in $K_{0.8}Fe_{1.7}Se_2$ have revealed a re-emergence of superconductivity around 48 K under a pressure of 12.5 GPa.[14] Superconductivity above 50 K has been reported in FeSe monolayer grown by molecular beam epitaxy (MBE).[15] Very recently, polycrystalline $A_xFe_{2-y}Se_2$ synthesized by liquid ammonia method shows a diamagnetic drop at temperatures of 30-46 K.[16]

The exciting developments naturally lead us to ask a basic question: Is it possible to obtain reproducible bulk superconductivity beyond 31 K in $A_xFe_{2-y}Se_2$ system by a conventional solid state reaction, just as the synthesis of other iron-based superconductors? If possible, it will be a crucial progress as we may stay away from toxic substances like liquid ammonia. Doubtlessly it will also greatly stimulate extensive and deep studies on the material due to the easily available synthesis condition in many groups. A related key question is how to determine the superconductivity if the new SC phase is small in size. These questions are experimentally answered in this paper. We demonstrate that by exactly controlling the amount of starting excess Fe we obtain a new phase other than the 30 K SC phase in $A_xFe_{2-y}Se_2$ crystal. We combine thermodynamics and ESR measurements to prove that the new phase becomes superconducting at 44 K. Structural analysis by energy-dispersive x-ray (EDX), Raman scattering and x-ray diffraction (XRD) indicate that the 44 K SC phase has a 122-like structure, but with an unexpectedly large c-axis lattice parameter of ~18 Å. By combination of the structural experiments and a comprehensive first-principles study, we reasonably deduce the structure of the new SC phase. Finally we find an interesting empirical law in FeSe-based systems, i.e., the square of maximum Tc is approximately proportional to the separation between neighbouring FeSe layers.

**Results**

The temperature dependences of resistivity and susceptibility are shown in Figs. 1a and 1b, respectively. A sharp drop in resistivity occurs around 44 K. And a clear kink in susceptibility is seen at the same temperature.(Inset of Fig. 1b) A full SC transition is observed at a lower temperature of ~30 K, which is obviously related to the 31 K SC phase mentioned above. The transition at 44 K is gradually suppressed by applied magnetic fields, which is very consistent with a SC transition. On the other hand, the transition does not demonstrate zero resistivity. In this sense, thermodynamic measurements alone are not enough to determine that it is a SC or just an antiferromagnetic (AF) transition. Meanwhile crystal characterization tells us the 44 K phase is only a small volume fraction of crystal. That severely limits the applications of most bulk techniques. ESR may be the most appropriate choice in this case as it has been proven to be extremely sensitive to SC transition.[16] It should be noted that much evidence for phase separation between 30K-SC and insulating AF phases has been accumulated so far, as mentioned above. But all the measurements in this paper cannot distinguish the subtle structural difference between the two phases. From the point of view of our measurements, the dominant 30K-phase mentioned hereafter, refers to the mixed phase of 30K-SC and its AF host ones.

Figure 2 shows ESR spectra from 1.8 to 300 K in the crystal with 44 K transitions.(Left panels) And the spectra from the crystal without 44 K anomaly are also shown here for comparison.(Central panel) Based on their relative signal levels, the spectra can be divided into three temperature ranges: 300~50 K, 45~35 K and 30~1.8 K. Resonant absorption is clearly observed in the first range. The ESR spectra in the two crystals are very similar both in the high-temperature range of 300 to 50 K and in the low-temperature range of 30 to 1.8 K. This similarity strongly suggests that the dominant phase in the two samples is the same. The resonance signal from free spins is gradually suppressed to zero with decreasing the temperature from 300 to 60 K. (Figs. 2a and 2d) The temperature-dependence of half-width of resonance peak, is shown in the insets of Figs. 2a and 2d. The evolution of the half-widths implies the disappearance of AF fluctuations around 60 K, consistent with previous work.[18] The sharp signal below 30 K at low fields typically reflects the magnetic shielding effect in a Meissner state. (Figs. 2c and 2f)[17] The disappearance of resonance signal may point to a weak magnetic transition, which may occur in the 30K-phase, or be extrinsically originated from magnetic impurities. However, the ESR measurements for all possible magnetic impurities, do not seem to support the extrinsic scenario.[36]

A key difference between ESR spectra in the two crystals appears in the intermediate temperature range. From 35 to 45 K, a strong absorption signal is seen in the crystal with 44 K phase. (Fig. 2b) In contrast, in the sample without 44 K phase no response is detected in the same temperature range. (Fig. 2e) This clearly indicates that the signal is contributed by the 44 K phase. There are three possible mechanisms for the absorption signal: antiferromagnetic resonance (AFMR), non-resonant absorption in a normal conducting state (NS) and in a superconducting state (SC). Let us discuss the mechanisms in more details. It looks the Lorentzian fitting based on AFMR works well (Fig. 2g). The fitting gives a constant resonance field and a width which increases with decreasing temperature (Insert of Fig. 2g), while a typical AFMR requires a strongly temperature-dependent resonance field due to the fast change of internal fields and a rapidly decreasing width with lowering temperature due to the drastic reduction of magnon-magnon and magnon-phonon scattering.[37-41] This excludes the possibility of AFMR. For a NS, the non-resonant microwave absorption is proportional to the surface resistance, which can be generally expressed as[19] $R_S = (\mu\omega\rho/2)^{1/2} = [\mu_0(1+\chi)\omega\rho/2]^{1/2}$, where $\rho$ and $\chi$ denote resistivity and susceptibility, $\mu_0$ and $\mu$ are the permeability of vacuum and medium, and $\omega$ microwave frequency. We use the field-dependent resistivity and relative susceptibility $\Delta\chi$ at 37 K (Figs. 1c and 1d), as inputs to the formula. Then we obtain a monotonic absorption curve, which completely fails to reproduce the dip feature, as shown in Fig. 2h. This unambiguously excludes a normal state as the origin of the special dip structure. While in a SC state, electromagnetic waves can induce very large instantaneous Josephson currents, which immediately destroy local superconductivity. Microwave absorption occurs in the transient process in a superconductor. And such a picture allows us to perfectly reproduce the absorption lineshape (Fig. 2i) (The detail can be found in the Supplementary).[20] Moreover, the critical field $H_{c1}$, taken at the spectral dip, also follows the conventional temperature-dependence of lower critical fields expected for a superconductor (Insert of Fig. 2i). These clearly demonstrate that the anomaly in resistivity and susceptibility at 44 K is a SC transition.

The strong signal of microwave absorption below 44 K (Fig. 2b) is also a crucial piece of evidence that the superconductivity at 44 K does not originate from filamentary phase. SEM measurements with a resolution of ~10 nm do not show any other phase except the above 30 an 44 K SC phases. That means filamentary phase must be less than 10 nm if existing. Such dimension is too small to form effective Josephson loops while a large amount of Josephson loops are necessarily required to output the observed strong signal.

In order to understand the structure of the new 44 K SC phase, we have further made careful structural measurements, as shown in Fig. 3. Consistent with thermodynamics and ESR measurements, X-ray diffraction clearly reveals there are two distinct phases in the sample showing 30 K and 44 K SC transitions.(Fig. 3a) Their c-axis lattice parameters are 14.11 Å and 18.10 Å, respectively. The former is the extensively studied 30 K SC phase with a vacancy-ordered structure.[3,10,12,24] The latter corresponds to the 44 K SC phase, which typically has a much longer c-axis as reported in the powder samples synthesized in liquid ammonia.[16]

Fig. 3b shows a SEM image of the crystal with the 44 K phase. SEM-EDX also displays the coexistence of two distinct phases: one has a K:Fe:Se ratio close to 0.8:1.7:2 (region II) and the other 0.9:2.17:2 (region I). Combined with the above measurements, consistently they are assigned to 30 K and 44 K SC phases. Fortunately, we are able to exactly pick up the same region I again in our microscopic Raman measurements. Raman spectra collected in Region II and I+II are shown in Fig. 3c. It is difficult to measure region I alone due to the size of the elliptical spot in a pseudo-backscattering configuration. Nevertheless, two different sets of Raman phonon modes can be explicitly separated by comparison. The Raman spectrum from region II exactly corresponds to the "$\sqrt{5}\times\sqrt{5}$" phase, which is ultimately connected with the superconductivity around 30 K.[27] The spectrum from 44 K SC phase (Region I) consists of four modes at 163, 188, 223, and 258 cm$^{-1}$. The four modes are actually the fingerprint of fully-filled FeSe layers, which has been confirmed by both first-principles calculations and experiment.[21-23,28]

**Discussion**

As we mentioned in Method, at this stage 44 K SC grains cannot be effectively separated from the dominant 30 K phase yet. It limits a direct resolution of the crystal structure by four-circle XRD. However, the consistent structural measurements have put strong constraints on the crystal structure of the new 44 K phase. EDX suggests ~20% excess Fe existing in the 44 K phase, which enter into proper sites to enlarge c-axis. Meanwhile Raman measurements tell us FeSe layers are fully filled rather than Fe-deficient and four-fold symmetry is preserved. Under symmetry and composition constraints, 2a, 2b, 4c and 4e positions are all the available symmetrical sites for accommodating excess Fe atoms. (Fig. 4) We do not consider the sites with lower symmetry like 8g because they will require many more atoms than the amount of 20% excess Fe.

We have carried out careful first-principles calculations to check the four cases. The structural parameters in the equilibrium structures are summarized in Table 1. Clearly 2a position is the first choice for excess Fe atoms. First, the c-axis lattice constant is the closest to the experimental one. Second, both the Fe-Se bond length and Fe-Se-Fe bond angle in this case show the smallest changes compared to original FeSe parent compounds.[29] It means the smallest distortion of FeSe layer, which is strictly required by Raman measurements. Finally, the maximum amount of excess Fe atoms required in this case is 50%, which most closely matches ~20% given by EDX. As comparison, for the 4c or 4e case the amount of excess Fe is 100% under full filling. Therefore, combining first-principles calculations with structural experiments, we can safely deduce that the new 44 K SC phase has a 122-like structure with excess Fe occupying 2a sites.

TABLE I: The equilibrium structural parameters of tetragonal cells when excess Fe(2) atoms locate at most possible Wyckoff positions with the I4/mmm group symmetry preserved. $d_{Fe-Se}$ is the Fe-Se bond length and $\alpha_{Fe-Se-Fe}$ the Fe-Se-Fe bond angle.

| Fe(2) position | a (Å) | c (Å) | $d_{Fe-Se}$ (Å) | $\alpha_{Fe-Se-Fe}$(°) |
|---|---|---|---|---|
| 2a | 3.87 | 19.20 | 2.32 | 72.0, 112.6 |
| 2b | 3.97 | 13.92 | 2.34 | 73.8, 116.2 |
| 4c | 4.76 | 9.43 | 2.51 | 84.0, 142.3 |
| 4e | 3.26 | 20.94 | 2.52 | 80.7, 132.5 |

The unusually large c-axis leads us to consider its relation to the maximum $T_c$. We have summarized the maximum $T_c$'s of the known FeSe-based superconductors at ambient pressure in Fig. 5, and plot their dependence on the distance between neighboring FeSe layers. Interestingly, we find that the maximum $T_c$ increases monotonically with the layer separation. A quantitative fit yields $T_c^{max} = 23\sqrt{(d-5.39)}$, where d is the interlayer separation in units of Å. This empirical formula implies that superconductivity in FeSe systems tends to favor two-dimensional FeSe structures. This result sheds further light on the search for higher $T_c$ values in FeSe-based compounds.

**Methods**

All samples were prepared by the Bridgman method. $Fe_{1+z}Se$ was first synthesized as a precursor by reacting Fe powder with Se powder at 750 °C for 20 h. K pieces and $Fe_{1+z}Se$ powder were put into an alumina crucible with nominal compositions $KFe_{2(1+z)}Se_2$. When setting z~0, we obtained the crystal with only a SC transition around 30 K.(The inset of Fig.2d) After many attempts, we grew the crystal with both 30 K and 44 K SC phases by exactly controlling z value to 0.37 ~ 0.38. The growth is completely reproducible with this ratio. The detailed growth procedure is similar to other FeSe-based superconductors and may be found elsewhere.[11,12] The resulting crystals were characterized by x-ray diffraction (XRD). Their elemental compositions were verified by scanning electron microscopy with energy-dispersive x-ray (SEM-EDX) analysis.

As structural measurements revealed, the islands of 44 K phases are scattered in the dominant 30 K phase and have a typical size of 60 μm measured by SEM-EDX. And there is no effective way to separate the two phases yet. It is an obstacle for further exploring the 44 K phase, but can be well overcome by a flexible combination of suitable bulk techniques, as we have done in the paper. The resistivity was measured by a standard four-probe method. The consistent resistivity measurements on different samples suggest that the superconductivity at 44 K does not come from surface or filamentary phase as in such phase resistivity would be expected to be fluctuating from sample to sample rather than consistent. The strong ESR signal below 44 K and SEM measurements also rules out filamentary phase as discussed in the text.

The DC magnetic susceptibility was measured under a range of magnetic fields. These measurements were performed in a physical property measurement system (PPMS, Quantum Design). Raman measurements were performed in a pseudo-backscattering configuration with a triple-grating monochromator (Jobin Yvon T64000), delivering a spectral resolution better than 0.6 cm$^{-1}$. The light source was a 532 nm solid-state laser (Torus 532, Laser Quantum) whose beam was focused into a spot of diameter approximately 60 microns on the sample surface. ESR measurements were performed using a Bruker EMX plus 10/12 CW-spectrometer at X-band frequencies (f = 9.36 GHz), equipped with a continuous He gas-flow cryostat to cover the temperature range 1.8-300 K.

To study the possible atomic positions for excess Fe, the first-principles electronic structure calculations were implemented by using the projector augmented wave method.[30-34] The exchange-correlation potentials were rep-resented by the generalized gradient approximation (GGA) with Perdew-Burke-Ernzerhof[35] formula. The energy cut-off for the plane waves was chosen to be 300 eV. A $6 \times 6 \times 1$ K-point mesh for the Brillouin zone sampling and the Gaussian smearing technique were adopted. The structure optimization made the forces on all relaxed atoms smaller than 0.03 eV/Å.

Acknowledgments

We thank Bruce Normand for reading the manuscript and X. H. Chen for helpful discussions, Z. M. Yu for assisting with numerical simulations and D. M. Wang for the EDX experiments. This work was supported by the NSF of China and the Ministry of Science and Technology of China (973 projects: 2011CBA00112 and 2012CB921701). A.M.Z was supported by the Fundamental Research Funds for the Central Universities, and the Research Funds of Renmin University of China (12XNH094).

**Author contributions**

Q.M.Z planed and conducted all the experiments and analysis. A.M.Z carried out the resistivity, magnetization, SEM, Raman measurements and analysis on ESR data, and prepared all the figures. T.L.X grew the crystals and made X-ray diffraction, resistivity and magnetization measurements. K.L made first-principles calculations. W.T and Z.R.Y made ESR measurements.

**Additional information**

Competing financial interests: The authors declare no competing financial interests.

FIG. 1: (a) Temperature-dependence of resistivity under different magnetic fields. (b) Temperature dependence of relative susceptibility under different fields. The inset indicates details around 44 K. (c) Field-dependence of resistivity at 37 K drawn from (a). (d) Field-dependence of relative susceptibility $\Delta\chi$ at 37 K drawn from (b). The reason for subtracting the susceptibility at 60 K is to eliminate the contribution from magnetic background. The data in (c) and (d) are carefully fitted to produce fitting functions (red lines), which are taken as input for the ESR fitting in a normal conducting state as described in the text. The detailed fitting functions and parameters can be found in the Supplementary.

FIG. 2: Left panel: ESR spectra of the crystal with a 44 K SC phase, divided into three temperature ranges: (a) 300 to 50 K, (b) 45 to 35 K and (c) 30 to 1.8 K. The central panel: (d), (e) and (f), show ESR spectra for the crystal without 44 K SC phase in the same temperature ranges. The sharp signal at zero field in (c) and (f) is an indication of Meissner state and the upward or downward curvature is caused simply by microwave measurement phase. Note that the signal level in (c) and (f) is one order of magnitude higher than that in (b). The right panel shows the fitting for 44K-phase using different absorption mechanisms: (g) Lorentzian fitting for antiferromagnetic resonance absorption (AFMR). The insert shows the temperature dependence of the fitting resonance fields and widths, which completely contradicts to a typical AMFR (See the text); (h) Non-resonant absorption in a normal conducting state (NS); (i) Non-resonant absorption in a superconducting state (SC). The insert shows that $H_{c1}$ taken at the spectral dip follows an approximate square-root behavior (red line).

FIG. 3: (a) Upper: X-ray diffraction pattern of the crystal containing both 30 K and 44 K SC phases. Lower: X-ray diffraction pattern of the crystal with only a 30 K phase. (b) SEM image of the crystal with

44 K SC phase. Dashed blue line indicates the border of Region I, in which EDX gives K:Fe:Se ratio of 0.9:2.17:2. The rest of the area is labeled as Region II, in which K:Fe:Se ratio is 0.8:1.7:2. The light green ellipse represents the laser spot in Raman measurements. (c) Comparison of Raman spectra. Red and blue lines show the spectra collected in Region I+II and in Region II alone, respectively. And the black one comes from the crystal with only a 30 K SC transition. The dashed lines mark the modes only contributed by Region I.

FIG. 4: Possible atomic positions of excess Fe(2) atoms (green balls) in equilibrium structures with I4/mmm group symmetry preserved. The excess Fe(2) atoms are in the (a) 2a, (b) 2b, (c) 4c and (d)4e, Wyckoff positions, respectively.

FIG. 5: Relation between maximum $T_c$ and the separation of neighbouring FeSe layers. Fitting the data gives an empirical law for $T_c$ (red line).

Figure 1

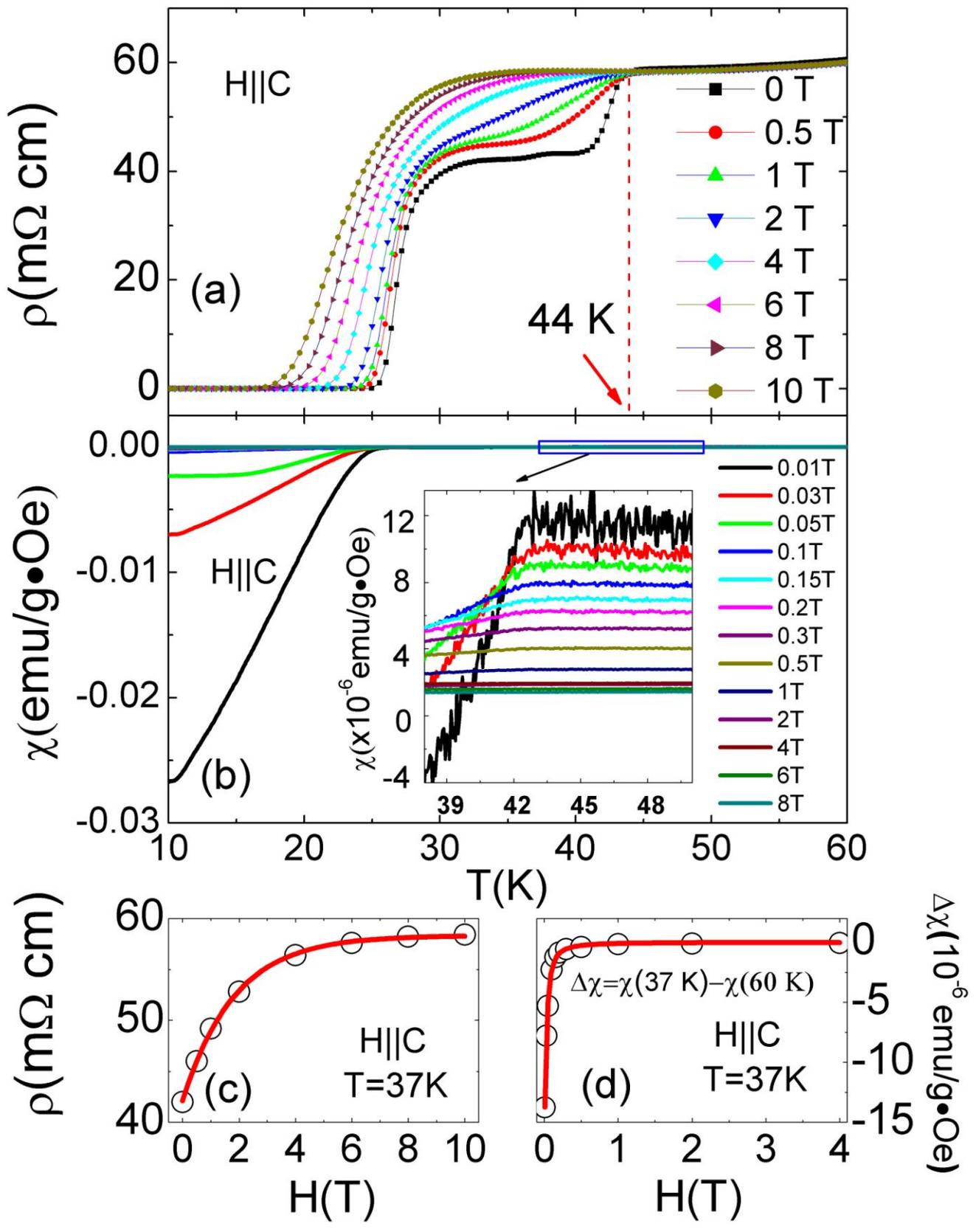

Figure 2

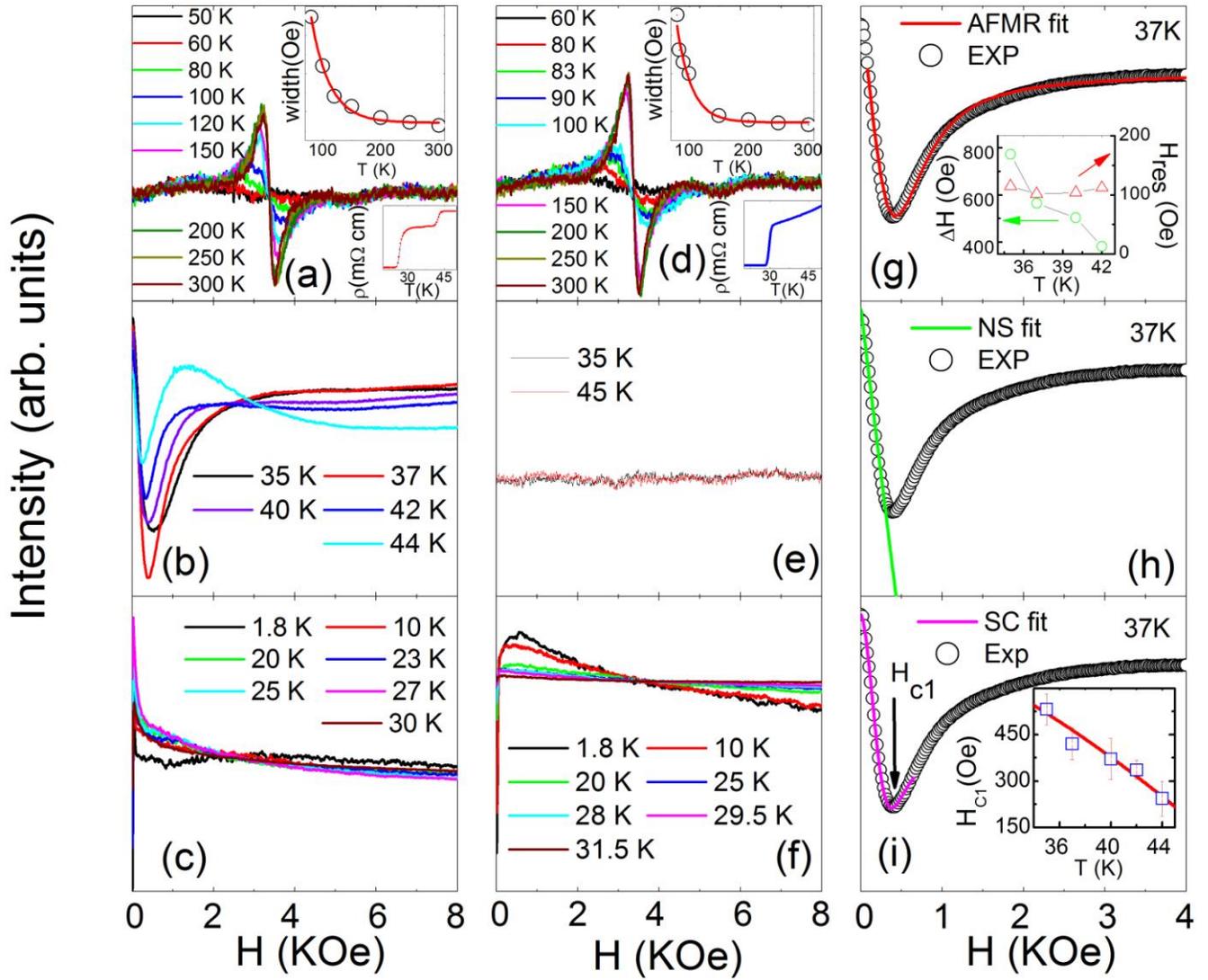

Figure 3

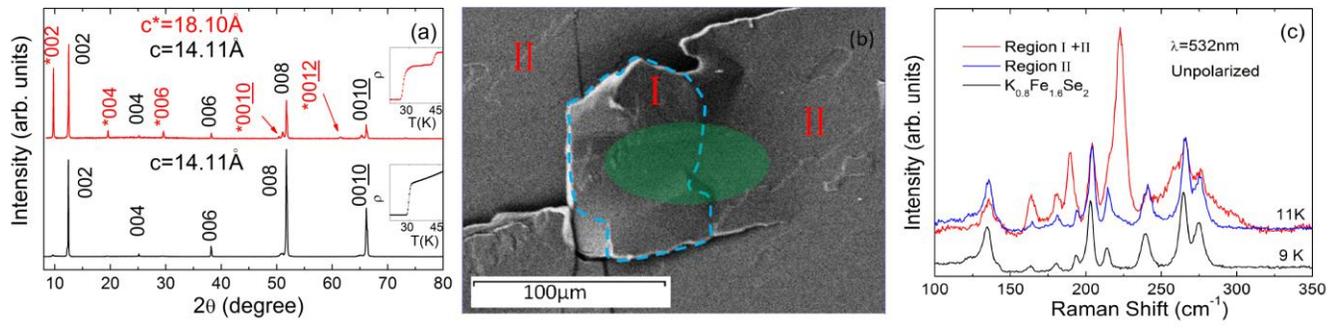

Figure 4

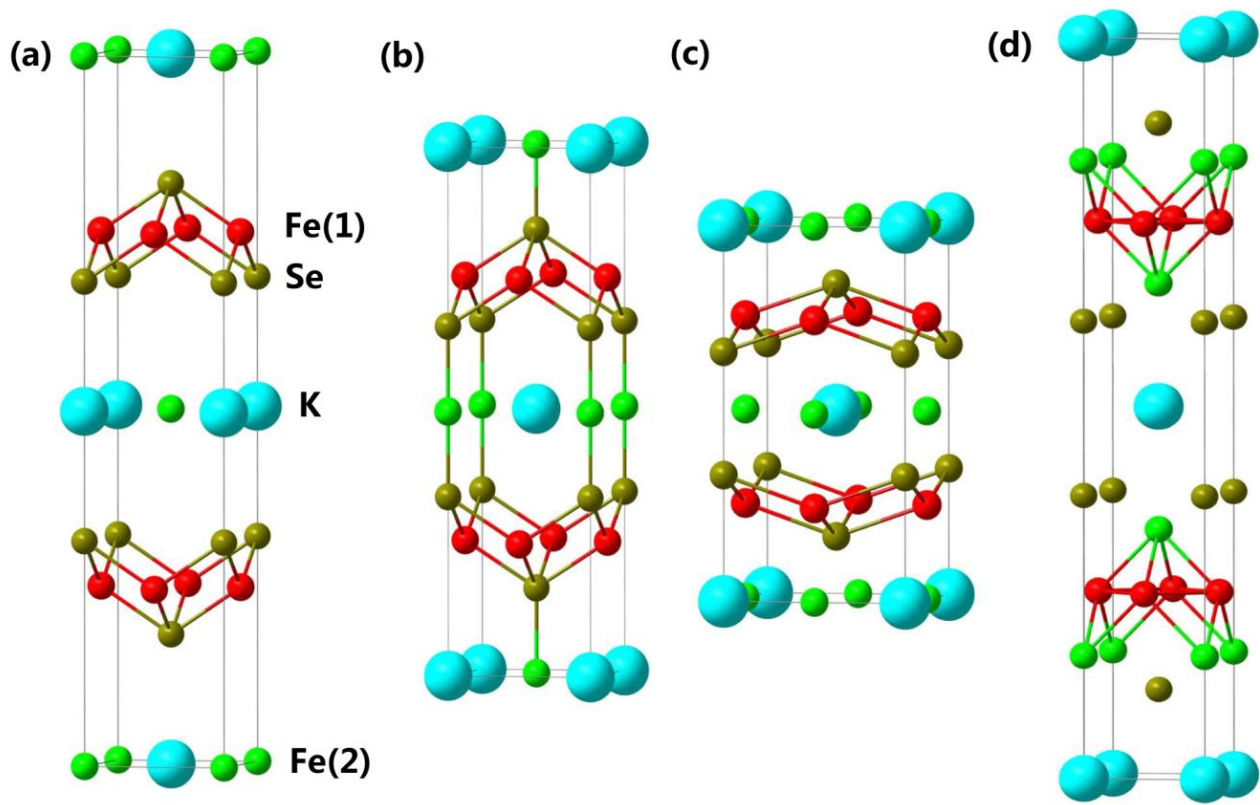

Figure 5

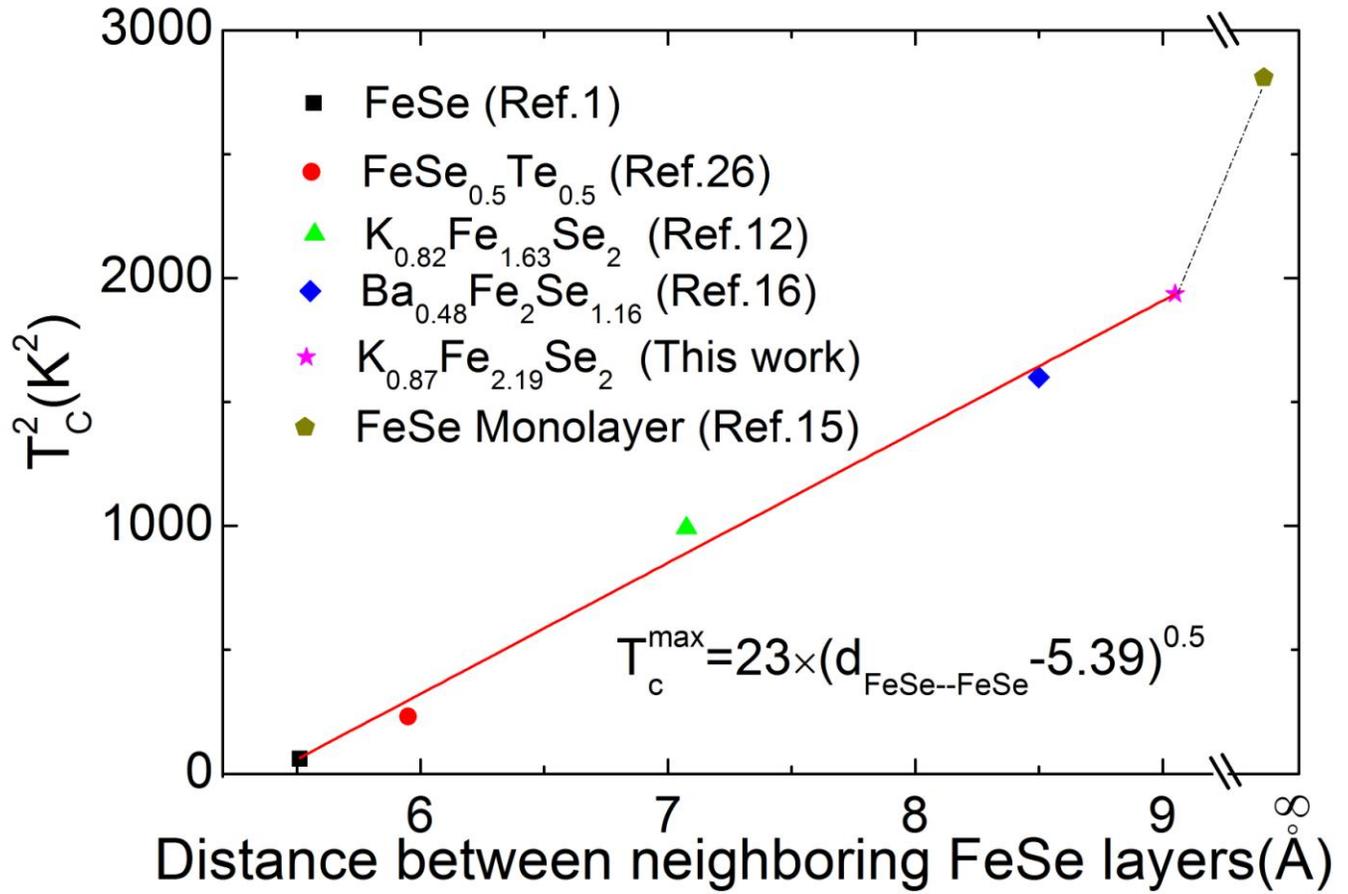

$T_c^{max} = 23 \times (d_{FeSe\text{--}FeSe} - 5.39)^{0.5}$

15